\documentstyle[prd,aps]{revtex}
\begin{document}
\draft
\preprint{OKHEP-00-03}
\title{\hfill OKHEP-00-03\\Comment on 
Casimir energy for spherical boundaries}
\author{I. Brevik\thanks{E-mail address:  iver.h.brevik@mtf.ntnu.no}}
\address{Division of Applied Mechanics, 
Norwegian University of Science and Technology,
N-7491 Trondheim, Norway}
\author{B. Jensen\thanks{E-mail address:  bjorn.jensen@hisf.no}}
\address{Faculty of Science, Sogn and Fjordane College,
N-6851 Sogndal, Norway}
\author{K. A. Milton\thanks{E-mail address: milton@mail.nhn.ou.edu}}
\address{Department of Physics and Astronomy,
The University of Oklahoma, Norman, OK 73019, USA}
\date{\today}

\maketitle
\begin{abstract}
It is shown that recent criticism by C. R. Hagen questioning the
validity of stress tensor treatments of the Casimir energy for
space divided into two parts by a spherical boundary is without foundation.
\end{abstract}


\section{Introduction}

In a recent paper, Hagen criticizes conventional, well-established
techniques for computing the Casimir stress on spherical, or in general,
curved boundaries \cite{hagen}.
In our view this paper is full of misconceptions and incorrect assertions,
and sheds no light on the subject.  The point seems to be that the
Green's function method for calculating the Casimir stress on a 
spherical shell, as originally proposed in Ref.~\cite{mds}, is incorrect, yet,
mysteriously, manages to arrive at the correct answer \cite{boyer}.  
The simpler,
and correct conclusion, would be, on the contrary, that that method
is valid.  As we will  demonstrate, such is the case.

Three errors are claimed to have been made in such calculations:
(i) the discontinuity of the
 stress tensor $T_{ij}$ has been used to calculate the force per unit area
on a spherical surface; (ii) poles are neglected in the rotation to
Euclidean space; and (iii) the incorrect outgoing wave boundary conditions
have been imposed on the Green's functions.  On the contrary, all 
of the procedures used in the criticized papers are, in fact, correct.

Let us first address the point (iii).  The causal or Feynman propagator is
used both in the Milton et al.\ papers \cite{mds,dm,mng,bm} 
and in Hagen's paper.  Hagen seems not to appreciate the simple identity obeyed 
by the free photon propagator:
\begin{eqnarray}
D_+({\bf r},t)&=&\int_{-\infty}^\infty {d\omega\over2\pi}
e^{-i\omega t}\int{(d{\bf k})\over(2\pi)^3}{e^{i{\bf k\cdot r}}\over
k^2-\omega^2-i\epsilon}
={i\over(2\pi)^2}{1\over r}
\int_{0}^\infty dk\,\sin kr \,e^{-ik|t|}
=\int_{-\infty}^\infty {d\omega\over2\pi}e^{-i\omega t}{1\over 4\pi r}
e^{i|\omega|r},
\label{causalprop}
\end{eqnarray}
where the second form is obtained by first integrating over $\omega$, the
third by integrating over $\bf k$. 
 The last form explicitly shows that
the casual propagator corresponds to outgoing-wave boundary 
conditions.\footnote{It is also
possible to use the retarded Green's function, which follows from a statistical
mechanics argument 
\cite{landau80,brevik99,dzyaloshinskii61,landau60,abrikosov75,%
barash75,lifshitz81}.  In general the connection
between the causal and the retarded Green's dyadics is
$\Im  \bbox{\Gamma}_+({\bf r}, {\bf r'},\omega)=
{\rm sgn}(\omega)\Im {\bf G}^R({\bf r},
{\bf r'},\omega).$  The same results are obtained.}
A pedagogical derivation of Eq.~(\ref{causalprop}) and other properties of
the causal propagator is given in Problem 31.9 in Ref.~\cite{ce}.

As to point (i):  There can be no doubt that the discontinuity of $T_{rr}$
across the spherical surface represents the stress per unit area exerted
on that surface. As shown in the Appendix of Milton
and Ng \cite{mng2}, this is equivalent to the
form given by the stationary principle of electrostatics, the force density
being ${\bf f}=-{1\over2}E^2\nabla\epsilon$
for a dielectric body.  (See for example, Ref.~\cite{ce}.)
Moreover, in most of Milton's papers, consistency was checked
explicitly by rederiving the formulae by at least two different methods,
either from the stress on the surface, or from the energy obtained
by integrating $T_{00}$ over all space.  In fact in the original paper 
\cite{mds}
a third independent variational method was used.  The stress tensor
formulation of the problem is further validated in Sec.~3.
The problem that Hagen addresses seems to be the result of using undefined
formal expressions [for example, the sum in his Eq.~(8) does not exist] so 
statements about coefficients vanishing in the limit of the enclosing sphere
becoming infinite are not meaningful, for one must regulate the expressions
before a limit can be taken.

The final objection (ii) concerns the existence of poles in the complex
frequency plane.  Indeed such poles occur in the lower half frequency plane
for the exterior contribution, when the exterior volume is infinite (no
large exterior sphere).  But exactly this issue was dealt with in the original
paper \cite{mds}, on p.~395,  
and will be explained in detail in the next section.

So none of Hagen's objections appear to be valid.  Of course, it
is not incorrect to enclose the sphere in question by a much larger sphere
to keep the eigenvalues real, but it is not necessary, and the
simplified procedure is correct.

There are many other minor misstatements and incomplete remarks that render 
Hagen's paper objectionable.  The recent 
experiments \cite{lamoreaux}, while admittedly based on
a somewhat different geometry than parallel plates, indeed confirm the theory
at the 5--1\% level because the correction to a spherical lens is easily
done. Moreover, there can be little room for scepticism about the reality
of Casimir forces, in view of the confirmation \cite{anderson}
of the closely related Lifshitz theory 
\cite{lifshitz,dzyaloshinskii61,landau60}, 
and the demonstration of the equivalence of van der Waals forces with
the Casimir force between and within dilute bodies \cite{brevik99,vdw}. 
 The Casimir force
for cylindrical and spherical geometries has recently been confirmed by
several authors using completely different zeta function techniques\cite{zeta}.
These, and other methods, also give finite results for other geometries
besides the ``electromagnetic sphere,'' namely for cylinders, for scalar
and fermionic modes, for dielectric bodies when $\epsilon\mu={\rm const}$,
etc.   We could go on, but our point is taken.

\section{On the legitimacy of rotating the integration contour}

Here we readdress in more detail 
the issue of ``rotating the contour of frequency
integration.''  As we will see, what is involved is a bit more sophisticated:
It is a Euclidean transformation.
The argument as given in \cite{mds} applied to the
Green's functions, either inside or outside the sphere of radius $a$, 
for which either
$r,r'<a$, or $r,r'>a$.  In fact, as employed in that reference and elsewhere,
the ``rotation'' is actually applied to the expression for the Casimir energy,
or the stress at the surface.  To perform the transformation, it is essential
that both inside and outside contributions be included.  We can, however,
consider the scalar (or TE) modes separately.  Those give
for the force per unit area on the sphere (where we have not
included a large external sphere---its inclusion does not modify the argument)
\begin{eqnarray}
f={i\over 8\pi^2a^3}\sum_{l=0}^\infty\left(l+{1\over2}\right)
\int_{C} d\omega \,e^{-i\omega
\tau}\left(ka\left[{H^{(1)\prime}_{l+1/2}(ka)\over H^{(1)}_{l+1/2}(ka)}
+{J'_{l+1/2}(ka)\over J_{l+1/2}(ka)}\right]+1\right),
\label{scalarcas}
\end{eqnarray}
where $k=|\omega|$.
(For the derivation of Eq.~(\ref{scalarcas}) see \cite{bm}.)
Here the sign of the additive constant has been reversed, in effect, by
adding a contact term, a term proportional to $\delta(\tau)$, so that
the resulting frequency integral be convergent.  Now it may be verified
that the integrand in Eq.~(\ref{scalarcas}) has the following analytic
properties in the complex variable $\zeta=ka$:
\begin{itemize}
\item The singularities lie in the lower half plane or on the real axis.
Consequently, the integration contour $C$ in $\omega$ lies just
above the real axis for
$\omega>0$, and just below the real axis for $\omega<0$.
\item For $\Im \zeta>0$,  the integrand goes to zero as $1/|\zeta|^2$.
(This is a weaker condition than specified in \cite{mds}.)  This convergent
behavior is the result of including both interior and exterior contributions.
\end{itemize}
Then we may write the stress on the sphere as
\begin{eqnarray}
f=\int_C{d\omega\over2\pi}e^{-i\omega\tau}g(|\omega|),
\label{stress}
\end{eqnarray}
 where the integrand satisfies the dispersion relation
 \begin{eqnarray}
 g(|\omega|)={1\over\pi i}\int_{-\infty}^\infty d\zeta{\zeta\over\zeta^2
 -\omega^2-i\epsilon}g(\zeta),
 \end{eqnarray}
 because the singularities of $g(\zeta)$ occur only for $\Im\zeta\le0$.
Now we can carry out the $\omega$ integral in Eq.~(\ref{stress}) to obtain
\begin{eqnarray}
f={1\over2\pi}\int_{-\infty}^\infty d\zeta{\zeta\over|\zeta|}e^{-i|\zeta||\tau|}
g(\zeta).
\end{eqnarray}
Finally, we re-write the result in Euclidean space by making the Euclidean
transformation
$i|\tau|\to|\tau_4|>0$,
so that we have the representation
\begin{eqnarray}
{1\over2|\zeta|}e^{-|\zeta||\tau_4|}=\int_{-\infty}^\infty {dk_4\over2\pi}
{e^{ik_4\tau_4}\over k_4^2+\zeta^2}.
\end{eqnarray}
Thus the Euclidean transform of the stress is
\begin{eqnarray}
f\to f_E=i\int{dk_4\over2\pi}e^{ik_4\tau_4}g(i|k_4|).
\end{eqnarray}
In effect, then, the Euclidean transformation is given by the recipe
$\omega\to ik_4$, $|\omega|\to i|k_4|$, $\tau\to i\tau_4$.
In particular, the force per unit area given for a massless scalar field
(\ref{scalarcas}) is transformed into the expression
\begin{eqnarray}
f_E=-{1\over8\pi^2 a^4}\sum_{l=0}^\infty(2l+1){1\over2}\int_{-\infty}^\infty
dy \,e^{i\delta y}\left(x{K'_{l+1/2}(x)\over K_{l+1/2}(x)}+x{I'_{l+1/2}(x)
\over I_{l+1/2}(x)}+1\right),
\end{eqnarray}
where we have adopted dimensionless variables,
$y=\omega a$, $\delta=\tau_4/a\to0$, $x=|y|$.
It is now straightforward to evaluate this expression.  The result is
as (surprisingly) first given in 1994 \cite{bm}:
$f=0.0028168/(4\pi a^4)$.
Thus a careful analysis shows that the methodology in Ref.~\cite{mds} is
correct, and that, in fact, as long recognized,
 a finite result is only achievable if the
difference between $T_{rr}(r=a-\epsilon)$ and $T_{rr}(r=a+\epsilon)$ is
taken.

\section{On the local interpretation of 
the stress tensor for a curved boundary}

\subsection{Theory}

Let us readdress the issue about the relation between the change
in the stress tensor $T^{\alpha\beta}$ 
across an {\em arbitrary} surface, and the resulting
force density $f^\alpha$ on the surface. In Hagen's paper it is argued
that the well-known relation between $f^\alpha$ and a change in
$T_{rr}$ across a planar surface in Cartesian coordinates is lost
when one is dealing with curved surfaces, and apparently even when
one is using non-Cartesian coordinates independent of the presence
or non-presence of curved surfaces. Hagen bases his reasoning on
the assumption that $\nabla_\alpha T^{\alpha\beta}=0$
holds everywhere. However, it is an almost trivial fact that this
is not the case when sources for the electromagnetic field are
present. It is in particular straightforward to show that one in
general has
\begin{equation}
\nabla_\alpha T^{\alpha\beta}=-F^{\beta\lambda}J_\lambda=-f^\beta
\end{equation}
where $J^\lambda$ is the source current, $F^{\alpha\beta}$ is
the electromagnetic field strength tensor, and $f^\beta$
is the four-vector force density acting on the material sources 
in the system.
Since this relation is covariant we conclude that the
interpretation of $f^\alpha$ as a force density holds in any
situation. This interpretation is in particular independent of the
explicit system of coordinates employed or whether one is dealing
with a curved spacetime. One can always locally recover the usual
Minkowski spacetime results by transforming to a local vierbein.

 The {\it total force} $F^\beta$ on the source is simply the
volume integral of the force density in some three-dimensional
spacelike hyper-surface $\Sigma$. The total force vector field on
a simply connected volume $V$ in $\Sigma$ with boundary $\partial
V$ is most simply computed using Stokes' theorem
\begin{equation}
F^\beta =\int_{V}f^\beta =-\int_{ V}\nabla_\alpha
T^{\alpha\beta}=-\oint_{\partial V} n_\alpha T^{\alpha\beta}\, .
\end{equation}
In this expression $n^\alpha$ represents a unit spacelike
outward normal to $V$. We will now specialize to the
case when we are dealing with a thin conducting spherical shell,
and when Maxwell's vacuum field equations hold everywhere away
from the shell. Thus $V$ represents the shell, and $\partial V$ its inner
and outer surfaces.
Clearly, in general the
total stress $F$ in the direction of a unit vector field
$u^\alpha$ is $F=\int_V f^\beta u_\beta$.
Due to time-independence and the exact spherical
symmetry of the shell, we choose $u^\alpha$ to point in the radial direction, 
and we find that
the total outward stress on the shell {\em per unit solid angle} is
\begin{equation}
{\cal F}=r_-^2 T^{rr}(r_-) -r_+^2 T^{rr}(r_+)\, .
\end{equation}
In this expression the components of the stress tensor are
to be evaluated at the inner and the outer radii of the shell,
which are denoted by $r_-$ and $r_+$ respectively. In the limit
when the shell is infinitely thin ($r_+\rightarrow r_-$) with
radius $a$ we thus find that the force per unit area is
\begin{equation}
f={{\cal F}\over a^2}=T^{rr}(a_-)-T^{rr}(a_+).
\label{stressdiff}
\end{equation}
The same
result can also be obtained by integrating the divergence of the
stress tensor directly, using the
fact that Tr$(T_{\alpha\beta})=0$.
This should prove once and for all that Hagen's critical
discussion about the use of the change in $T_{rr}$ as a direct
measure of the stress on a spherical shell is, from the theoretical
side, completely wrong. Let us now turn to the experimental side.

\subsection{Experiments}

  We wish to present 
two counterexamples from optics which demonstrate that Hagen's claim 
about the use of $T_{rr}$ in calculating the force on a spherical shell is 
incorrect in this regard: The predictions obtained from local stress tensor 
calculations are in accordance with experimental evidence.

1.  The first example concerns the oscillations of a small water droplet 
illuminated by a laser pulse. In 1988 Zhang and Chang 
\cite{zhang88} took photographs of 
water droplets (50 $\mu$m radius) at various instants after illumination by 
a $\lambda$=0.60 ${\mu}$m dye laser pulse of 0.4 ${\mu}$s duration. In the 
case of a 100 mJ pulse, the droplet began oscillating with surface elevations 
of up to 30 per cent of the equilibrium radius. Since a spherical drop acts as 
a lens, the maximum elevations were observed, as expected,  at the rear section.

From a theoretical point of view this problem is a hybrid problem: the natural 
way to proceed is to calculate the electromagnetic surface force density using 
Eq.~(\ref{stressdiff}) as 
if the surface of the sphere were at rest. Thereafter one can 
calculate the hydrodynamic response of the sphere by means of the 
Navier-Stokes equation. The problem was studied by Lai {\it et al.} 
\cite{lai89}, and by Brevik and Kluge \cite{brevik99a},  for various 
polarizations in the linear approximation for the surface elevation. 
Wave optics was required. Although the linear approximation may appear 
rough at the rear section, it is clear from the figures that this kind of 
theory is in reasonable agreement with the observations. This example, as 
far as we are aware, is one of the very few cases where the observations of 
the radiation pressure are {\it local}. The viability of the expression 
(\ref{stressdiff}) is quite evident.

2.  Our second example is about the optical forces observed on microparticles 
in an evanescent laser field. There is a striking experimental demonstration 
of this effect, by Kawata and Sugiura \cite{kawata92}: The particles are 
lifted up from the dielectric interface and moved parallel to the interface. 
Although this effect is not a local effect such as above, one has 
nevertheless, in order to describe it theoretically, to make use of the same 
formula (\ref{stressdiff}) 
as before. The theory of the effect has been worked out by Almaas 
and Brevik \cite{almaas95}, Lester and Nieto-Vesperinas \cite{lester99}, and 
others. The agreement between wave theory and observations is reasonable, 
again justifying the legitimacy of Eq.~(\ref{stressdiff}).

\section*{Acknowledgements}
The work of KAM was supported in part by a grant from the US Department of
Energy.  We thank Ricardo Moritz Cavalcanti for detecting a typographical
error in an earlier version of this paper.

\end{document}